\documentclass[9pt,twocolumn,twoside]{osajnl}
\journal{ol} % Choose journal (ao,jocn,josaa,josab,ol,optica,pr)

\setboolean{shortarticle}{true}
\usepackage{lineno}
\usepackage{graphicx}
\usepackage{siunitx}
\usepackage{amsmath}
\usepackage{url}

\title{Satellite-assisted laser magnetometry with mesospheric sodium}

\author[1]{Tong Dang}
\author[2,3,*]{Emmanuel Klinger}
\author[2,3,4]{Felipe Pedreros Bustos}
\author[2,3]{Arne Wickenbrock}
\author[5]{Ronald Holzl{\"o}hner}
\author[2,3,6]{Dmitry Budker}

\affil[1]{Department of Electrical and Systems Engineering, University of Pennsylvania, Philadelphia, PA 19104, USA}
\affil[2]{Johannes Gutenberg-Universit{\"a}t Mainz, 55128 Mainz, Germany}
\affil[3]{Helmholtz-Institut Mainz, GSI Helmholtzzentrum f{\"u}r Schwerionenforschung, 55128 Mainz, Germany}

\affil[4]{Laboratoire d’Astrophysique de Marseille, Aix Marseille Universit\'e, UMR 7326 CNRS, CNES, 13013 Marseille, France}
\affil[5]{European Southern Observatory (ESO), 85748 Garching bei München, Germany}

\affil[6]{Department of Physics, University of California, Berkeley, California 94720, USA}

\affil[*]{Corresponding author: eklinger@uni-mainz.de} %% 

\begin{abstract}
Magnetic field sensing provides crucial insights into various geophysical phenomena such as atmospheric currents, crustal magnetism, and oceanic circulation. In this paper, a method for remote detection of magnetic fields using mesospheric sodium with an assisting satellite is proposed. Sodium atoms in the mesosphere are optically pumped with a ground-based laser beam. A satellite-borne detector is used to measure magneto-optical rotation of the polarization of a probe laser beam by the sodium atoms. This sensitive magnetometry method benefits from direct detection of laser photons and complements existing space- and aircraft-borne techniques by probing magnetic fields at upper-atmospheric altitudes inaccessible to those.
\end{abstract}

\setboolean{displaycopyright}{true}

\begin{document}

\maketitle

\section{Introduction}
While a ground-based approach for remote magnetometry has been developed by detecting the sodium atomic fluorescence at resonance \cite{bustos2018,PedrerosBustos:18b}, a broad angular distribution of the emitted fluorescence limits the flux of photons that can be detected at the ground telescope. One idea to overcome this limitation is to use mirrorless lasing by the sodium atoms in the mesosphere \cite{Rui2021}, but this concept has yet to be implemented in on-sky experiments. Besides mirrorless lasing, an alternative approach is to employ an assisting satellite with a photodetector or a retroreflector, so that the polarization and the intensity of transmitted probe light can be measured directly, obtaining stronger signals. However, this configuration poses new challenges associated with the motion of the satellite. 

Thanks to advancing technologies, small and inexpensive satellite platforms such as CubeSat have gained popularity since the 2000s. Hundreds of CubeSats were launched in the past decade due to their cost-effective components and availability of launch opportunities \cite{spacenews,centennial1}. A standard CubeSat unit (1U) is $10\times10\times11$\,cm$^3$ and weights $1.33$\,kg or less \cite{CubeSat}, with the 3U CubeSat being the most popular as its larger volume allows more components to be incorporated \cite{Polat2016}. Despite its small size, a CubeSat can feature attitude control and detumble systems such as reaction wheels and magnetorquers \cite{lu2018,Farissi2019}. Additionally, optical communication and high-speed data transmission have been demonstrated via optical up- and downlinks, allowing opportunities for real-time measurements \cite{mata-calvo2017,osborn2021,mata-calvo2019}. CubeSat missions have become increasingly capable and versatile since the first proposal in 1999, making direct measurements of transmitted light possible for remote sky magnetometry. 

\begin{figure}[!htbp]
    \centering
    \includegraphics[width=0.49\textwidth]{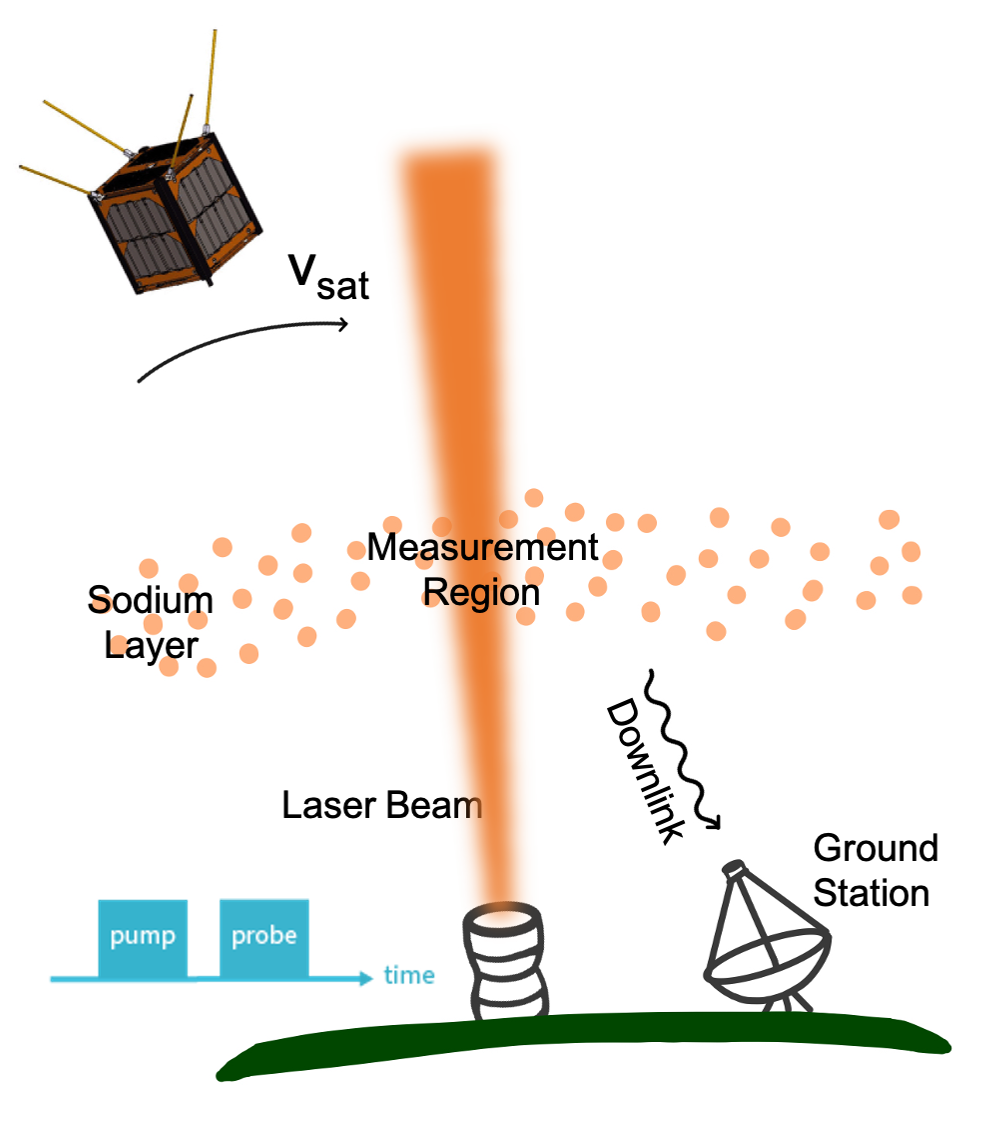}
    \caption{A schematic illustration of satellite-assisted remote magnetometry. The inset shows an example of the control sequence of the laser beam.}
    \label{fig:schematic1}
\end{figure}

Building upon these advances, the new proposed system involves a laser that serves as both the pumping and probing beams as well as a trackable miniaturized satellite. A satellite that can be optically tracked is launched into low earth orbit (LEO) at an altitude of 500\,km and is equipped with a photodetector and a polarizer nearly orthogonal to the laser polarization. The laser beam first optically pumps mesospheric sodium in the target-measurement region, as the satellite orbits through the measurement region, the laser beam probes the sodium atoms that have been optically pumped. Photodetectors on the satellite detect the polarization of the transmitted light. The nonlinear magneto-optical rotation (NMOR) has a time-dependent component at the Larmor frequency ($\Omega_L$) or its second harmonic, with $\Omega_L$ proportional to the magnetic field strength. A schematic of the configuration is shown in Fig.\,\ref{fig:schematic1}.

\section{Measurement procedure} \label{sec:2}
\begin{figure}[!htbp]
    \centering
    \includegraphics[width=0.49\textwidth]{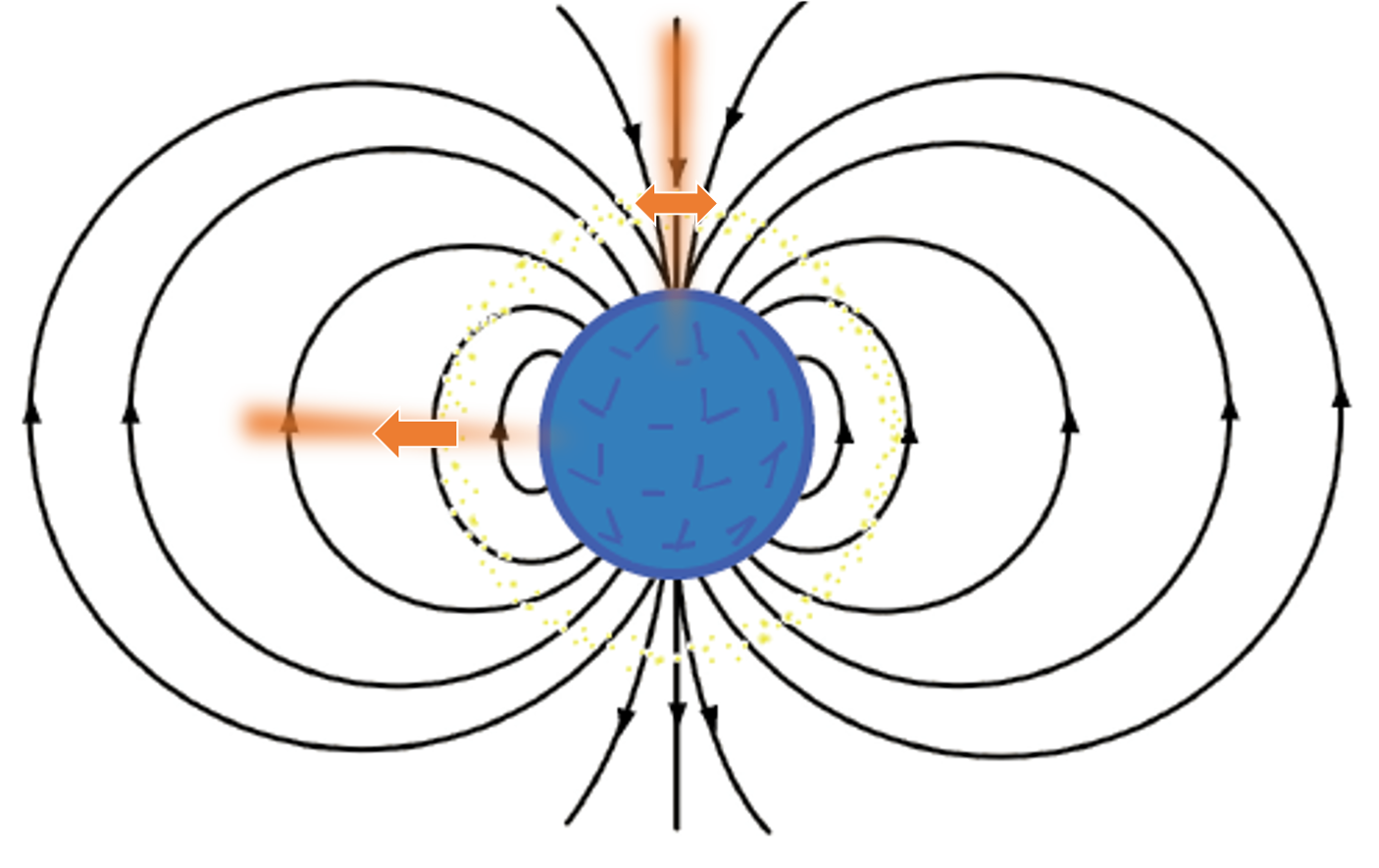}
    \caption{Measurements of the Earth magnetic field near the poles and at the equator. Orange arrows show atomic polarization from optical pumping relative to the magnetic field.}
    \label{fig:schematic2}
\end{figure}

After optical pumping, the Larmor frequency will be measured by detecting the magneto-optical rotation ($\varphi_1$) and/or the transmission ($T$) of the polarized probe-laser light. In general, the geomagnetic field makes an arbitrary angle with the laser beam. However, one can identify two extreme cases, namely measurements near the poles and at the equator, respectively, as shown in Fig.\,\ref{fig:schematic2} where the analysis is particularly simple. Here, the magnetic field lines run, correspondingly, parallel or perpendicular to the light beam. We emphasize that this technique can be used at arbitrary latitudes. 

Near the poles, we use linearly polarized laser light for pumping, which creates atomic alignment (see, for example, Ref.\,\cite{auzinsh2010}). 
% Due to the precession of the polarization in the geomagnetic field, synchronous optical pumping needs to be utilized.
The linearly polarized probe laser light will show Faraday rotation of $\varphi_1$ that can be measured with a polarimeter on board the satellite. The analyzer can be configured with a transmission polarizer oriented nearly orthogonal ($\varphi_0$) to the polarization of the light. The intensity at the detection port is then
\begin{equation}
    I_{\text det} \approx I_{0}\sin^2(\varphi_0+\varphi_1),
    \label{eqn:Idet}
\end{equation}
where $I_0$ is the light irradiance in the mesosphere.

An estimate of $\varphi_1$ near the poles can be obtained from 
\begin{equation}
\varphi_{1P} \approx \frac{\alpha}{2}\cdot\sin(2g_F\mu_B B t + \psi)\cdot e^{-\gamma_0 t},
\label{eq:phi1P}
\end{equation}
where $\alpha$ is approximately 4\% given by the optical absorption  of the sodium layer, $g_F$ is the Landé $g$-factor for ground state Na, $\mu_B$ is the Bohr magneton, $\psi$ is the optical pumping phase, and $\gamma_0$ is the spin relaxation rate in the mesosphere of typically about $1/(250\,\mu\text{s})$ \cite{bustos2018}.

At the equator, the laser beam would be nearly perpendicular to the magnetic field of the Earth. Optical pumping can then be achieved with circularly polarized light, and magnetometry is performed by either measuring $\varphi_1$ with linearly polarized probing light, or by measuring the transmission of a circularly polarized probing laser beam. Optical rotation at the equator can be estimated as 
\begin{equation}
    \varphi_{1E} \approx \frac{\alpha}{2}\cdot\sin(g_F\mu_B B t + \psi)\cdot e^{-\gamma_0 t}.
    \label{eq:phi1E}
\end{equation}
In either case $\varphi_1 \ll \varphi_0 \ll 1$, and \eqref{eqn:Idet} can be reduced into
\begin{equation}
    I_{det} = I_{0}\sin^2(\varphi_0+\varphi_1) \approx I_0{(\varphi_0+\varphi_1)}^2 \approx I_{0}({\varphi_0}^2 + 2\varphi_0\varphi_1).
    \label{eqn:Idet2}
\end{equation}

If we are limited by photon shot noise [see Sec.\,\ref{sec:3}], the sensitivity of the polarimeter does not depend on the angle between the polarizer and analyzer $\varphi_0$. This is because the useful signal [see \eqref{eqn:Idet2}] is bilinear in the rotation angle in sodium $\varphi_1$ and $\varphi_0$, while the photon shot noise is dominated by the fluctuations of the leading term proportional to the square root of $\varphi_0^2$. If $I_0$ itself fluctuates, for example due to intensity noise, light scattering in clouds, etc., these fluctuations will contribute to noise in proportion to $\varphi_0^2$. It is thus beneficial to keep $\varphi_0$ as small as possible. However, this angle should be much larger than the angle corresponding to finite extinction of the polarizer and analyzer. In conjunction with this, we note that the laser depolarization on the uplink is expected to be negligible \cite{collet1972, toyoshima2009} and that the value of $\varphi_0$ can be adjusted at the launch telescope.

Another approach in this scenario is to investigate the transmission spectrum of circularly polarized probe light. Although the measurements are scalar, vector information can be obtained as the NMOR resonances depend on the magnetic field direction \cite{pustelny2006}. In general, the signal from the probe light contains first and second harmonics of the Larmor frequency, whose ratio can give information about the direction of the magnetic field. For linearly polarized light with frequency modulated at $\Omega_m$, the main resonance occurs at $\Omega_m = 2\Omega_L$ when the magnetic field is along the light propagation direction, as in the case of measurements near the poles. For measurements near the magnetic equator in which the magnetic fields may be tilted in the plane perpendicular to the direction of light propagation, an additional resonance appears at $\Omega_m = \Omega_L$. The amplitudes of the resonance corresponds to the tilt of the magnetic field relative to the light propagation direction, allowing quantitative determination of the magnetic field direction.

The procedure for the experiment could be as follows. Once the satellite is within the laser spot, pumping is performed for about a millisecond (several ground-state spin relaxation times). The pump beam could be modulated (in amplitude, frequency or polarization) synchronously with the Larmor precession (see \cite{bustos2018,PedrerosBustos:18b} and references therein) or it could be a strong continuous light pulse. The pumping interval is followed by a probing interval of comparable duration during which polarization rotation is monitored. Because spins are locked by the pump beam (both in the case of cw and synchronous pumping), the relative phase between pumping and probing is the same for all probed sub-layers of the mesosphere. For this reason, we expect propagation effects in the 10\,km-long layer to be negligible. Note that for the case where pump light is circularly polarized and the probe is linearly polarized, the light polarization must be switched between pumping and probing.  After a measurement is complete, the cycle can be repeated as long as the satellite is within the laser-beam. An attractive option is to utilize a tracking telescope \cite{chang2019}, which would then allow magnetic measurements along the entire arc traced by the satellite while it is visible by the ground observer.  

\section{Sensitivity estimates: photon shot noise} \label{sec:3}

For state-of-the-art remote magnetometry, i.e. fluorescence based mesospheric Na magnetometry, the fundamental sensitivity is hard to achieve because only a tiny fraction of fluorescent photons is detected within the small solid angle subtended by ground-based detectors. In this case, the sensitivity is limited by photon shot noise \cite{higbie2011}. Likewise, the shot noise of the detected photons is a limiting factor in satellite-borne detection. The photon shot noise of an ideal polarimeter is solely determined by the total number of measured photons $N_p$, such that
\begin{equation}
 \delta   \varphi =\frac{1}{2\sqrt{N_p}}\equiv\frac{1}{2\sqrt{\phi_{Sat} \tau}},
\end{equation}
where $\tau$ is the measurement time and $\phi_{Sat}$ is the photon flux picked up by the satellite. 

A typical guidestar laser emits a power of 20\,W at a wavelength of 589\,nm (orange), which corresponds to a flux of $\phi_0\approx 6\times10^{19}$ photons per second. The emitted beam has a typical divergence due to diffraction of
\begin{equation}
    \delta\theta = \frac{1.22\lambda}{D_{GT}} \approx \SI{2.4\,}{\micro\radian},
\end{equation}
where $\lambda= 589\,$nm is the wavelength of the light resonant with the sodium D line and $D_{GT}=\SI{30}{cm}$ is the diameter of the guidestar launch telescope projecting the laser beam onto the sky \cite{RonBeamSimulation2008}. To choose the operation parameters, one has to account for the pumping/probing irradiances in the sodium layer. The pump beam power has to be sufficiently high for the pumping to be efficient, while the power of the probe beam should be low enough to prevent repolarization. The ($1/e^2$) beam diameter at the sodium layer ($H_{\text Na}=90-100$\,km) is $d_{\text Na} = 2 \,\delta\theta \,H_{\text Na} \approx \SI{0.5}{\m},$ due to diffraction alone and roughly twice this value in the presence of turbulence. 

The dimensionless optical pumping saturation parameter for the mesospheric sodium atoms can be estimated as in, for example, 
Ref.\,\cite{OptPolAtoms}, Sec.\,9.1, taking into account that the effective ground-state relaxation is dominated by the Larmor precession in the magnetic field of the Earth at a frequency of a few hundred kHz. With full light power and the assumed spot size, this saturation parameter is on the order of $10^2 \gg 1$. This means there is more than enough light to polarize the sodium atoms, but on the other hand, the probe beam power would need to be attenuated (or frequency-detuned), which would, in turn, adversely affect the photon shot noise (signal size). However, if the area of the beam in the mesosphere was to be increased by two orders of magnitude, the repolarization effect would be minimized without affecting the pumping efficiency. We thus choose the beam diameter at the sodium layer as 5\,m corresponding to approximately $24\,\rm{m}$ at an altitude of $H_{\text sat}=500$\,km. At this altitude, the orbital speed is given by
\begin{equation}
    v_{\text sat} = \sqrt{\frac{G\,M_E}{R_E+H_{\text sat}}} \approx 7610\, \rm{m\,s^{-1}},
\end{equation}
where $G = 6.67408 \times 10^{-11} {\text m}^3 \,{\text kg}^{-1} \,{\text s}^{-2}$ is the gravitational constant and $M_E$ and $R_E$ are the mass and radius of the Earth, respectively. With a measurement shot lasting about 2\,ms, averaging over 300 shots retains horizontal spatial resolution of one  kilometer (we assume that the satellite is being tracked while the field is measured in the sodium layer below it, along its path).

Because of the limited size of the CubeSat, only a fraction of the total photon flux $\phi_0$ will be detected. Assuming an on-board detector with a diameter of $D_{SD}=10$\,cm, the fraction of the detected photon flux is about $1.7\times10^{-5}$ leading to $\phi_{Sat}\approx 1.0\times 10^{15}\,\rm{s^{-1}}$. With these numbers, we estimate a photon shot-noise limit of $\delta  \varphi \approx 4.8 \times 10^{-7}\,\rm{rad}$ per single shot ($\tau =1\,$ms). From \eqref{eq:phi1E}, a crude estimate of the shot-noise-limited magnetic sensitivity for $t\sim 1/\gamma_0$ reads
\begin{equation}
    \delta B_E \sim \frac{11\gamma_0}{\alpha g_F \mu_B}\,\delta\varphi_{1E},
\end{equation}
which leads to $\delta B_E\sim 77\,\rm{pT}$ (per single shot measurement of about $2$\,ms duration) at the equator. At the poles, $\delta B_P = \delta B_E/2 \sim 39\,\rm{pT}$ because the signal oscillates at twice the Larmor frequency, as discussed in Sec.\,\ref{sec:2}. 

With a bandwidth of $1/4\tau=250\,$Hz, accounting for the $\sim 1\,$ms pumping time, the conversion to a sensitivity per root Hz yields $4.9\,\rm{pT}/\sqrt{\rm{Hz}}$ when measuring at the equator, and half that value at the poles. This is about two orders of magnitude better compared to sensitivities derived for fluorescence-based sky magnetometry \cite{higbie2011}.

\section{Conclusions and Outlook}
We have presented a concept for magnetometric measurements in the upper atmosphere based on magneto-optical rotation in the mesospheric sodium layer. A laser beam resonant with an atomic sodium transition is launched with a ground-based laser and detected using a polarimeter on board a small satellite. Sensitivity estimates show that this technique, based exclusively on existing well-proven technologies, is highly promising and will complement the existing measurement techniques based on the currently employed balloons and aircraft on the one hand, and space-borne instruments on the other, addressing the range of altitudes of considerable current interest to geophysics \cite{matzka2021,laundal2021}. Going beyond the estimates presented here, one should perform detailed modeling of the experiment \cite{LGSBloch}, taking into account factors like the magnetic field gradients across the sodium layer, the effect of the wind and the satellite-tracking dynamics, etc., to optimize the operation parameters (the exact timing pattern, laser-frequency tuning with respect to the atomic resonance, hyperfine repumping, etc.), as well as considering various avenues for achieving better performance. For example, we have assumed a relatively small area of 100\,cm$^2$ of the satellite-borne detector which results in wasting a lot of photons and thus raising shot noise. The light collection may be improved by several orders of magnitude by deploying an origami-folded collection mirror in orbit along the lines of well-proven space technologies \cite{ikeya2020significance,miyazaki2018deployable}. Finally, we note that the proposed method can be used also for other types of measurements in addition to magnetometry. For example, one can extract the density of sodium atoms integrated over the mesospheric layer from the amplitude of optical rotation signal; measurements of electric field may also be possible \cite{budker2002nonlinear}.

\section*{Acknowledgements}
We are indebted to Dr. Alexander Akulshin (Swinburne), Dr. Justin Albert (UVIC), Dr. Domenico Bonaccini Calia (ESO), Dr. Trevor Bowen (SSL), Prof. Paul Hickson (UBC), Dr. Frank Lison (TOPTICA Photonics), and Dr. Noelia Martinez (ANU) for most useful discussions. This work was supported in part by the German Federal Ministry of Education and Research (BMBF) within the Quantentechnologien program (FKZ 13N15064). FPB received support from the European Union’s Horizon 2020 research and innovation programme under the Marie Sk\l{}odowska-Curie grant agreement No. 893150.

\bibliography{bibliography}

\begin{thebibliography}{10}
\newcommand{\enquote}[1]{``#1''}

\bibitem{bustos2018}
F.~{Pedreros Bustos}, D.~{Bonaccini Calia}, D.~Budker, M.~Centrone,
  J.~Hellemeier, P.~Hickson, R.~Holzl{\"o}hner, and S.~Rochester,
  {\protect\JournalTitle{Nature communications}} \textbf{9}, 1 (2018).

\bibitem{PedrerosBustos:18b}
F.~{Pedreros Bustos}, D.~{Bonaccini Calia}, D.~Budker, M.~Centrone,
  J.~Hellemeier, P.~Hickson, R.~Holzl\"{o}hner, and S.~Rochester,
  {\protect\JournalTitle{Opt. Lett.}} \textbf{43}, 5825 (2018).

\bibitem{Rui2021}
R.~Zhang, E.~Klinger, F.~Pedreros~Bustos, A.~Akulshin, H.~Guo, A.~Wickenbrock,
  and D.~Budker, {\protect\JournalTitle{Phys. Rev. Lett.}} \textbf{127}, 173605
  (2021).

\bibitem{spacenews}
D.~Leone, \enquote{First booz allen satellite will observe air force laser,}
  \url{https://spacenews.com/first-booz-allen-satellite-will-observe-air-force-laser/}
  (2015).

\bibitem{centennial1}
{Gunter's Space Page}, \enquote{Centennial 1,}
  \url{https://space.skyrocket.de/doc_sdat/centennial-1.htm}.

\bibitem{CubeSat}
{The CubeSat Program, Cal Poly SLO}, \enquote{{CubeSat} design specification,}
  \url{https://static1.squarespace.com/static/5418c831e4b0fa4ecac1bacd/t/56e9b62337013b6c063a655a/1458157095454/cds_rev13_final2.pdf}.

\bibitem{Polat2016}
H.~C. Polat, J.~Virgili-Llop, and M.~Romano, {\protect\JournalTitle{Journal of
  small satellites}} \textbf{5}, 513 (2016).

\bibitem{lu2018}
W.-C. Lu, L.~Duan, and Y.-X. Cai, \enquote{De-tumbling control of a {CubeSat},}
  in \emph{2018 IEEE International Conference on Advanced Manufacturing
  (ICAM),}  (IEEE, 2018), pp. 298--301.

\bibitem{Farissi2019}
M.~S. Farissi, S.~Carletta, A.~Nascetti, and P.~Teofilatto,
  {\protect\JournalTitle{Aerospace}} \textbf{6}, 133 (2019).

\bibitem{mata-calvo2017}
R.~Mata-Calvo, D.~B. Calia, R.~Barrios, M.~Centrone, D.~Giggenbach,
  G.~Lombardi, P.~Becker, and I.~Zayer, \enquote{Laser guide stars for optical
  free-space communications,} in \emph{Free-Space Laser Communication and
  Atmospheric Propagation XXIX,} , vol. 10096 (International Society for Optics
  and Photonics, 2017), p. 100960R.

\bibitem{osborn2021}
J.~Osborn, M.~J. Townson, O.~J. Farley, A.~Reeves, and R.~M. Calvo,
  {\protect\JournalTitle{Optics Express}} \textbf{29}, 6113 (2021).

\bibitem{mata-calvo2019}
R.~M. Calvo, J.~Poliak, J.~Surof, A.~Reeves, M.~Richerzhagen, H.~F. Kelemu,
  R.~Barrios, C.~Carrizo, R.~Wolf, F.~Rein \emph{et~al.}, \enquote{Optical
  technologies for very high throughput satellite communications,} in
  \emph{Free-Space Laser Communications XXXI,} , vol. 10910 (International
  Society for Optics and Photonics, 2019), p. 109100W.

\bibitem{auzinsh2010}
M.~Auzinsh, D.~Budker, and S.~Rochester, \emph{Optically polarized atoms:
  understanding light-atom interactions} (Oxford University Press, 2010).

\bibitem{collet1972}
E.~Collett and R.~Alferness, {\protect\JournalTitle{JOSA}} \textbf{62}, 529
  (1972).

\bibitem{toyoshima2009}
M.~Toyoshima, H.~Takenaka, Y.~Shoji, Y.~Takayama, Y.~Koyama, and H.~Kunimori,
  {\protect\JournalTitle{Optics express}} \textbf{17}, 22333 (2009).

\bibitem{pustelny2006}
S.~Pustelny, W.~Gawlik, S.~Rochester, D.~J. Kimball, V.~Yashchuk, and
  D.~Budker, {\protect\JournalTitle{Physical Review A}} \textbf{74}, 063420
  (2006).

\bibitem{chang2019}
J.~Chang, C.~Schieler, K.~Riesing, J.~Burnside, K.~Aquino, and B.~Robinson,
  \enquote{Body pointing, acquisition and tracking for small satellite laser
  communication,} in \emph{Free-Space Laser Communications XXXI,} , vol. 10910
  (International Society for Optics and Photonics, 2019), p. 109100P.

\bibitem{higbie2011}
J.~M. Higbie, S.~M. Rochester, B.~Patton, R.~Holzl{\"o}hner, D.~B. Calia, and
  D.~Budker, {\protect\JournalTitle{Proceedings of the National Academy of
  Sciences}} \textbf{108}, 3522 (2011).

\bibitem{RonBeamSimulation2008}
R.~{Holzl{\"o}hner}, D.~{Bonaccini Calia}, and W.~{Hackenberg},
  {\protect\JournalTitle{Proc. SPIE}} \textbf{7015}, 701521 (2008).

\bibitem{OptPolAtoms}
M.~Auzinsh, D.~Budker, and S.~Rochester, \emph{Optically polarized atoms:
  understanding light-atom interactions} (Oxford University Press, 2010).

\bibitem{matzka2021}
J.~Matzka, C.~Stolle, Y.~Yamazaki, O.~Bronkalla, and A.~Morschhauser,
  {\protect\JournalTitle{Space Weather}} \textbf{19}, e2020SW002641 (2021).

\bibitem{laundal2021}
K.~Laundal, J.-H. Yee, V.~G. Merkin, J.~W. Gjerloev, H.~Vanham{\"a}ki, J.~P.
  Reistad, M.~Madelaire, K.~Sorathia, and P.~Espy,
  {\protect\JournalTitle{Journal of Geophysical Research: Space Physics}} p.
  e2020JA028644 (2021).

\bibitem{LGSBloch}
{LGSBloch Mathematica} package, available at
  \url{http://rochesterscientific.com/ADM/}.

\bibitem{ikeya2020significance}
K.~Ikeya, H.~Sakamoto, H.~Nakanishi, H.~Furuya, T.~Tomura, R.~Ide, R.~Iijima,
  Y.~Iwasaki, K.~Ohno, K.~Omoto \emph{et~al.}, {\protect\JournalTitle{Acta
  Astronautica}} \textbf{173}, 363 (2020).

\bibitem{miyazaki2018deployable}
Y.~Miyazaki, {\protect\JournalTitle{Proceedings of the IEEE}} \textbf{106}, 471
  (2018).

\bibitem{budker2002nonlinear}
D.~Budker, D.~Kimball, S.~Rochester, and V.~Yashchuk,
  {\protect\JournalTitle{Physical Review A}} \textbf{65}, 033401 (2002).

\end{thebibliography}

\end{document}